\documentclass[aps,prd,epsfig,preprintnumbers,superscriptaddress,10pt]{revtex4}

\usepackage{graphicx}
\usepackage{dcolumn}
\usepackage{bm}

\newcommand{\be}{\begin{equation}}
\newcommand{\ee}{\end{equation}}
\newcommand{\bea}{\begin{eqnarray}}
\newcommand{\eea}{\end{eqnarray}}
\newcommand{\ba}{\begin{eqnarray}}
\newcommand{\ea}{\end{eqnarray}}

\newcommand{\beq}{\begin{equation}}
\newcommand{\eeq}{\end{equation}}
\newcommand{\beqa}{\begin{eqnarray}}
\newcommand{\eeqa}{\end{eqnarray}}
\newcommand{\beqar}{\begin{eqnarray*}}
\newcommand{\eeqar}{\end{eqnarray*}}

\def\x{{\bf x}}

\begin{document}

\preprint{CERN-PH-TH/2011-264}

\title{Out of Medium Fragmentation from Long-Lived Jet Showers}

\author{Jorge Casalderrey-Solana} 
\affiliation{Departament d'Estructura i Constituents
de la Mat\`eria, 
Universitat de Barcelona, Mart\'\i \ i Franqu\`es 1, 08028 Barcelona, Spain}
\affiliation{Physics Department, Theory Unit, CERN, CH-1211 Gen\`eve 23, Switzerland}
    
\author{Jos\'e Guilherme Milhano}
\affiliation{CENTRA, Instituto Superior T\'ecnico, Universidade T\'ecnica de Lisboa, Av. Rovisco Pais, P-1049-001 Lisboa, Portugal}
\affiliation{Physics Department, Theory Unit, CERN, CH-1211 Gen\`eve 23, Switzerland}

\author{Paloma Quiroga Arias}
\affiliation{ LPTHE, UPMC Univ. Paris 6 and CNRR UMR7589, Paris, France}


\begin{abstract}
We study the time structure of vacuum jet evolution via a simple uncertainty principle estimate in the kinematic range 
explored by current heavy ion collisions at the LHC. We observe that a large fraction of the partonic splittings occur at 
large times, of the order of several fm. We compare the time distribution of vacuum splittings with the distribution of path lengths
traversed by jets in a heavy ion collision. We find that if no medium induced modification of the jet dynamics were present, a very large fraction
(larger than 80\% for inclusive jets) of the jet splittings would occur outside of the medium.  We confront this observation with current
available data on jet properties in heavy ion collisions and discuss its implications for the dynamics of jet-medium interactions.
\end{abstract}

\maketitle

\section{Introduction}

Jets  have for long been regarded as one of the main diagnostic tools for the study of the hot and dense matter produced in ultrarelativistic heavy ion collisions. The modification of jet observables in those collisions, as compared to their vacuum  counterpart (proton-proton collisions) encodes detailed information on the properties of the created medium. However, the unambiguous extraction of these  properties poses
challenges, both theoretical and experimental, which significantly complicate these studies.

On the theoretical side, a detailed understanding of the interaction between the parton showers and the QCD medium is required. There is a broad consensus on the pivotal role of medium induced gluon radiation as the main dynamical mechanism responsible for this interplay (see for example \cite{Wiedemann:2009sh,d'Enterria:2009am,CasalderreySolana:2007zz,Majumder:2010qh,Jacobs:2004qv,Gyulassy:2003mc} for recent reviews). However, different model implementations of this mechanism for the phenomenological description of single particle suppression, both at RHIC \cite{Adler:2002xw,Adler:2003qi} and the LHC \cite{Aamodt:2010jd},
 lead to the extraction of rather diverse medium parameters \cite{Armesto:2011ht}. Moreover, in recent years the importance of other dynamical mechanism, such as collisional energy loss \cite{Wicks:2005gt,Schenke:2009ik}, interference effects \cite{MehtarTani:2010ma,MehtarTani:2011tz,Armesto:2011ir,CasalderreySolana:2011rz}, non-trivial modification of the in medium fragmentation pattern \cite{Beraudo:2011bh}, AdS/CFT inspired mechanism (see \cite{CasalderreySolana:2011us} for a review), etc, has been stressed; this body of work illustrates the difficulties that models based solely on radiative losses face 
when confronted to data.

On the experimental side, jet measurements in a heavy ion environment lead also to many challenges mainly due to the large hadronic activity in the collision. However, modern jet reconstruction algorithms
\cite{Cacciari:2008gp,fastjet_fast} have led  to the development of strategies to take into account
both the large background and its fluctuations \cite{Cacciari:2010te,Aad:2010bu,Chatrchyan:2011sx}, and hence rendering those measurements possible. In fact, 
jet measurements carried out during the first LHC heavy ion run  \cite{Aad:2010bu,Chatrchyan:2011sx,Angerami:2011is,Collaboration:2011nsb} present serious challenges to current model formulations. On the one hand, jets that traverse a significant medium length lose a considerable amount of energy. Remarkably, this energy loss is accompanied by a very mild, at most, jet deflection \cite{Aad:2010bu,Chatrchyan:2011sx}, thus precluding an origin in a dynamical mechanism where a significant enhancement of radiation of semi-hard gluons is of importance. To accommodate these data, a simple mechanism based on the early decorrelation of soft jet components has been proposed by some of the authors \cite{CasalderreySolana:2010eh} (see \cite{Qin:2010mn,Young:2011qx,He:2011pd} for other descriptions of data based on radiative losses).
On the other hand, preliminary measurements of  jet fragmentation functions coincide, within errors,  with those obtained in p-p collisions.
The apparent contradiction between the unmodified fragmentation functions and the  large distortion of di-jet asymmetry distributions may point towards an underestimate of the effect  of the fluctuating background, as pointed out in \cite{Cacciari:2011tm}. However, if as claimed by both ATLAS and CMS these fluctuations are under control, the measurements show, 
in a nutshell, that although jets loose a significant fraction of their energy while traversing the medium, their fragmentation pattern is unchanged. These observations seem, {\it a priori}, at odds with the assumption of medium induced gluon radiation as the main mechanism responsible for energy degradation of in-medium jets.

Motivated by this experimental observation, we carry out here a simple study of the time structure of the fragmentation process of jets in vacuum. Since little is known about the details of the space-time dynamics of this evolution, we will rely, as customary in Monte Carlo implementations of in-medium jet dynamics \cite{Armesto:2009zc,Armesto:2009ab,Zapp:2008gi,Renk:2010zx}, on uncertainty principle arguments to determine the characteristic emission time of the different partons produced during hadronization. The goal of this note is to determine the typical time scales for the $vacuum$  emission of fragments in high energy jets within the kinematic range of the heavy ion experiments at the LHC. As we will see, for those jets the partonic showers develop at times of several  fm. 
Remarkably, this is larger than the typical path length that jets traverse through matter in one such collision.

 The paper is organized as follows: In section \ref{vactime}, the time structure of vacuum QCD evolution is studied. Since we are motivated by the properties of fragmentation functions, we will focus on the time distribution of partonic splittings and determine the fraction of splittings that occur at time larger than a given value $L$. In section \ref{geometry}, we determine the path length distribution of jets in heavy ion collisions. To take into account surface biases, a simple absorption (quenching) model for jets is used and the path length distributions for leading and associated jets are obtained. In section \ref{foutside}
 we convolute both distributions to determine the fraction of splittings that in an {\it unmodified jet} would occur outside the medium created in heavy ion collisions. We conclude the paper in section  \ref{discussion}
 by discussing how our vacuum jet analysis can help constraining models for in-medium modification of jets.

\section{\label{vactime}Estimating the time structure of QCD vacuum showers.}

The description of final state QCD branching is a perfect example of the predictive power of perturbative QCD.
While both the evolution equations that dictate the branching dynamics and their implementation in modern Monte Carlo event generators have become standard textbook material (see for example \cite{Ellis:1991qj}), they are invariably formulated in momentum space and little is known about the space-time structure of the branching process. 
Most attempts to understand this structure rely on an uncertainty principle argument to estimate the typical emission times of partons:
in a time-like parton shower, the typical lifetime of a parton of virtuality $Q$ (i.e. the time elapsed prior to its splitting into two less virtual objects) is given, in its rest frame, by $\tau_f\sim 1/Q$. Thus, a parton with energy $E$ in the centre of mass frame of the collision has its lifetime boosted to
\be
\label{tau_f}
\tau_f=2 \frac{E}{Q^2}\, ,
\ee
where the factor 2 ensures that for the emission of soft gluons with four momenta $(\omega, k_\perp, k)$, the formation time coincides with the usual relation $\tau_f=2 \omega/k^2_\perp$ . 
In this note we assume that the above argument holds, generalizing it to the estimation of the fragmentation times of partons in a QCD shower.

Eq.~(\ref{tau_f}) describes the lifetime of a single virtual excitation. However, final state partons in a typical high energy QCD process undergo several perturbatively describable splittings with the decay time of a given parton in the chain dictated by the kinematics of the fragmenting parton at that point in the chain.  Hence, eq.~(\ref{tau_f}) determines the time to emission relative to the parton being formed. Since DGLAP evolution imposes a strict ordering in virtuality, with those splittings involving higher virtuality occurring earlier, the time for splitting of a parton following $n$ previous splittings is given by
\be
\label{taus}
\tau_S=\sum^{n}_{i=1} \tau_f(i)\, .
\ee

These assumptions allow us, provided the virtuality at each step of the decay chain is known, to estimate the typical times needed for the evolution of a vacuum QCD shower. 
We note that the determination of the relevant virtualities is necessarily model dependent. Although solidly grounded in QCD, the practical implementation of the evolution process in event generators requires modeling assumptions that go beyond the controlled approximations in which QCD calculations are performed. Even within a given event generator several different schemes for the branching process are considered. The aim of this note is to provide a simple, yet realistic, estimate. As such we resort to a specific scheme implemented in PYTHIA \cite{Sjostrand:2006za}  where the evolution variable is the mass of the splitting virtual object and energy momentum conservation is imposed at each splitting step. Within this implementation we can reconstruct the branching chain of hard patrons in any process and assign an emission $\tau_S$ to each  splitting.

In all event generators, DGLAP evolution is carried down to a minimal scale $Q_0$ after which the dynamics is 
non-perturbative. Since different event generators vary considerably at this stage we will refrain from discussing the time structure at such late stages. 
However, in order to account for the dynamics at this stage we will add, for each fragment, a final splitting to the perturbative chain at a time given by eq. (\ref{tau_f}) with virtuality $Q=Q_0$ ($Q_0= 1$ GeV in PYTHIA). This time can be understood as the lifetime of a parton prior to its fragmentation onto hadrons.

Of particular interest for the purposes of this note is the probability of splitting \textit{after} a given time $\tau$. 
To construct this probability, and for reasons that will become apparent in the next section, we will perform a longitudinal boost from the collision centre of mass frame to the frame where the parent parton is transverse to the beam direction. In the remainder of the paper we will refer to the time in  this boosted frame simply as time.

\begin{figure}
\centering
\includegraphics[angle=0,width=.55\linewidth]{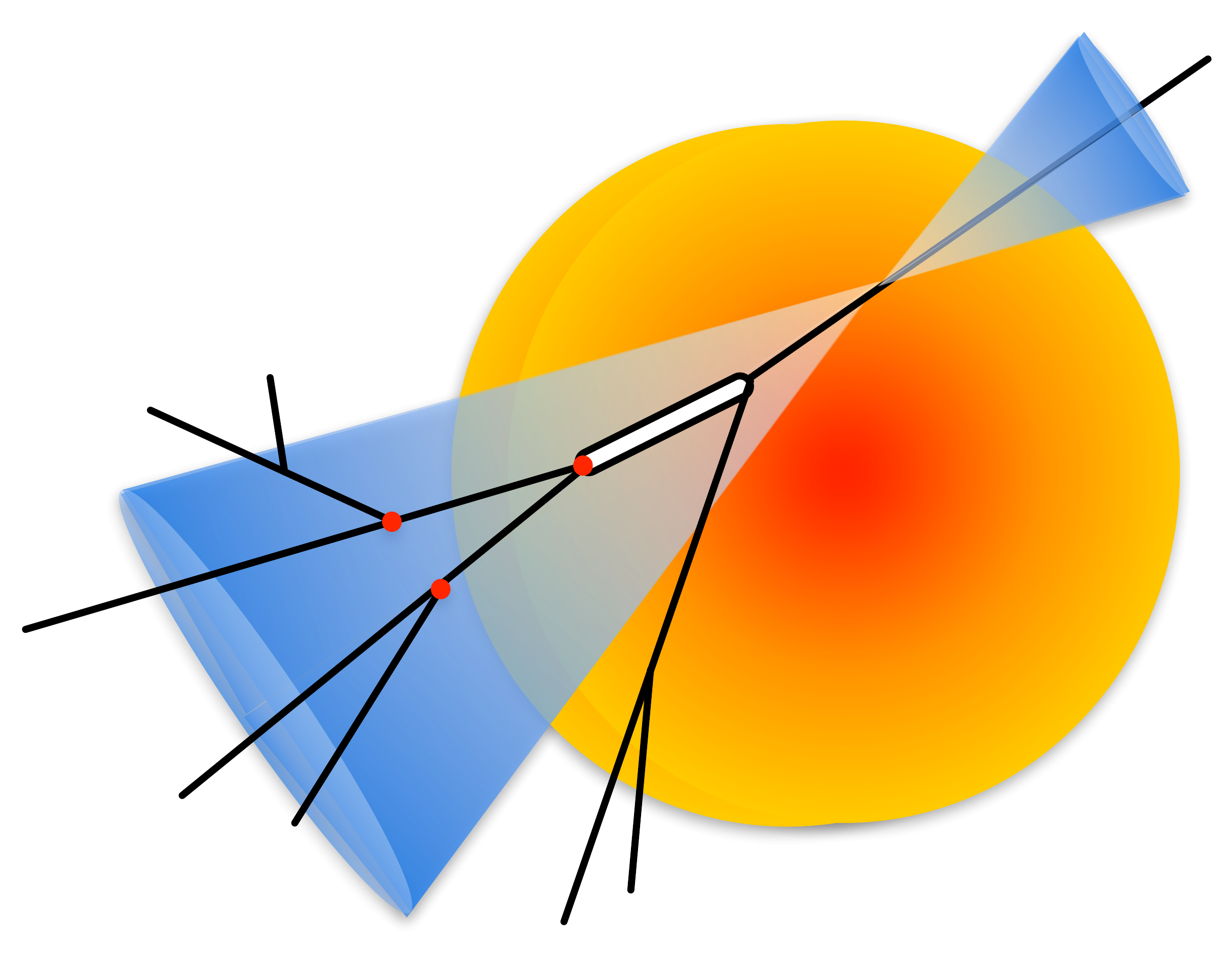}
\caption{\label{dibuxino}
Sketch of the evolution process for the associated jet in a di-jet event. The fragmentation proceeds by several  splittings
from the initial hard vertex  up to the final particles. The jet reconstruction procedure, represented by the blue 
cone, does not include all the fragments generated in the evolution. Thus, not all the splittings are directly 
connected to the particles used for the 
jet reconstruction. Note that all the fragments within the cone originate from a common ancestor (the parent parton, marked with a double line) and prior splittings  only change the overall kinematics of  that parton. 
From all the splittings in the chain, only $n_R$ splittings, represented by circles, are seeded by the parent parton and connected to the final reconstructed fragments, these splittings influence the most the final fragment 
distribution (up to kinematical changes).}
\end{figure}

The emission time of a given step in the decay chain depends on prior splittings, see eq.~(\ref{taus}). Thus, information on the chain back to the hard vertex is required. However,
fragmentation functions are mostly sensitive to the dynamics of the closest common ancestor (parent parton)
to the final fragments and branchings prior to the parent parton only result in overall kinematical changes.
Thus, for each final parton, we reconstruct the full chain of branchings back to the parent parton. To avoid double counting in jets with more than one final state parton,  splittings common to two or more particles are counted only once. Also, only splittings that are causally connected to those partons that fall within the specific jet selection and reconstruction procedure will be taken into account. Fig. \ref{dibuxino} provides a pictorial description of this procedure.

The probability distribution  $\mathcal{P}^{out}_R(L)$ for splitting after a given time $L$ is obtained from the event generator as the average over the Monte Carlo sample 
\be
\mathcal{P}^{out}_R(L)=\left<\frac{n^{out}_R(L)}{n_R}\right>\, ,
\ee 
where the index $R$ indicates that only branchings in the chains of those partons that end up within the jet definition are included,  $n_R$ is the total number of such splittings in a an event, and $n^{out}_R(L)$ the number of those that occur at $\tau_S>L$. The distribution  $\mathcal{P}^{out}_R(L)$  depends on the specific procedure employed to select and reconstruct the jet sample, in particular on the jet definition, via the Monte Carlo average.

As we have stressed repeatedly, the above procedure provides only an estimate of the typical distribution since it is
based on the identification of the typical scale of the emission process and assumes that all the splittings occur at such fixed time. This procedure clearly overestimates the emission time, since in reality it is given by a distribution in times with characteristic value $\tau_f$, with  emission both at earlier and later times than $\tau_f$ being allowed. To account for this spread,  we will use a simple assumption for the emission time distribution: the probability that a given splitting with typical formation time $\tau_f$ occurs at time $\tau$, $D(\tau)$ is given by
\be
\label{randtau}
D(\tau)=\frac{1}{\tau_f}e^{-\tau/\tau_f}\, .
\ee
This allows us to assign to each splitting a (random) emission time and, by following the discussion around eq.~(\ref{taus}),
determine the absolute time of each emission.  We will use these two evaluations of the emission time --- with a fixed typical time given by eq.~(\ref{tau_f}) and with a random sampling around $\tau_f$, eq.~(\ref{randtau}) --- as an estimate of the uncertainty in the distribution.

We constructed our event samples by generating di-jet events with PYTHIA 6.4~\cite{Sjostrand:2006za}  for  
pp collisions at a centre-of-mass energy of 2.76 TeV without underlying event.  
Jets were reconstructed at the partonic level 
using the anti-$k_t$ sequential recombination algorithm~\cite{Cacciari:2008gp} for different values of R, as implemented in FastJet~\cite{fastjet_fast}.
The events were selected with criteria based on those used by the CMS collaboration~\cite{CMS-analysis-summary}. A minimum $p_{T,leading}$ of 100 GeV was required for the leading jet. Once such jet was found, the subleading jet was required to have $p_{T,associated}>$40 GeV and to be sufficiently separated in azimuth from the leading jet, $\Delta\phi_{1,2}\ge 2\pi/3$. Further,  a cut in rapidity was imposed, selecting only jets within $|y|<$2.

\begin{figure}
\centering
\includegraphics[angle=0,width=.49\linewidth]{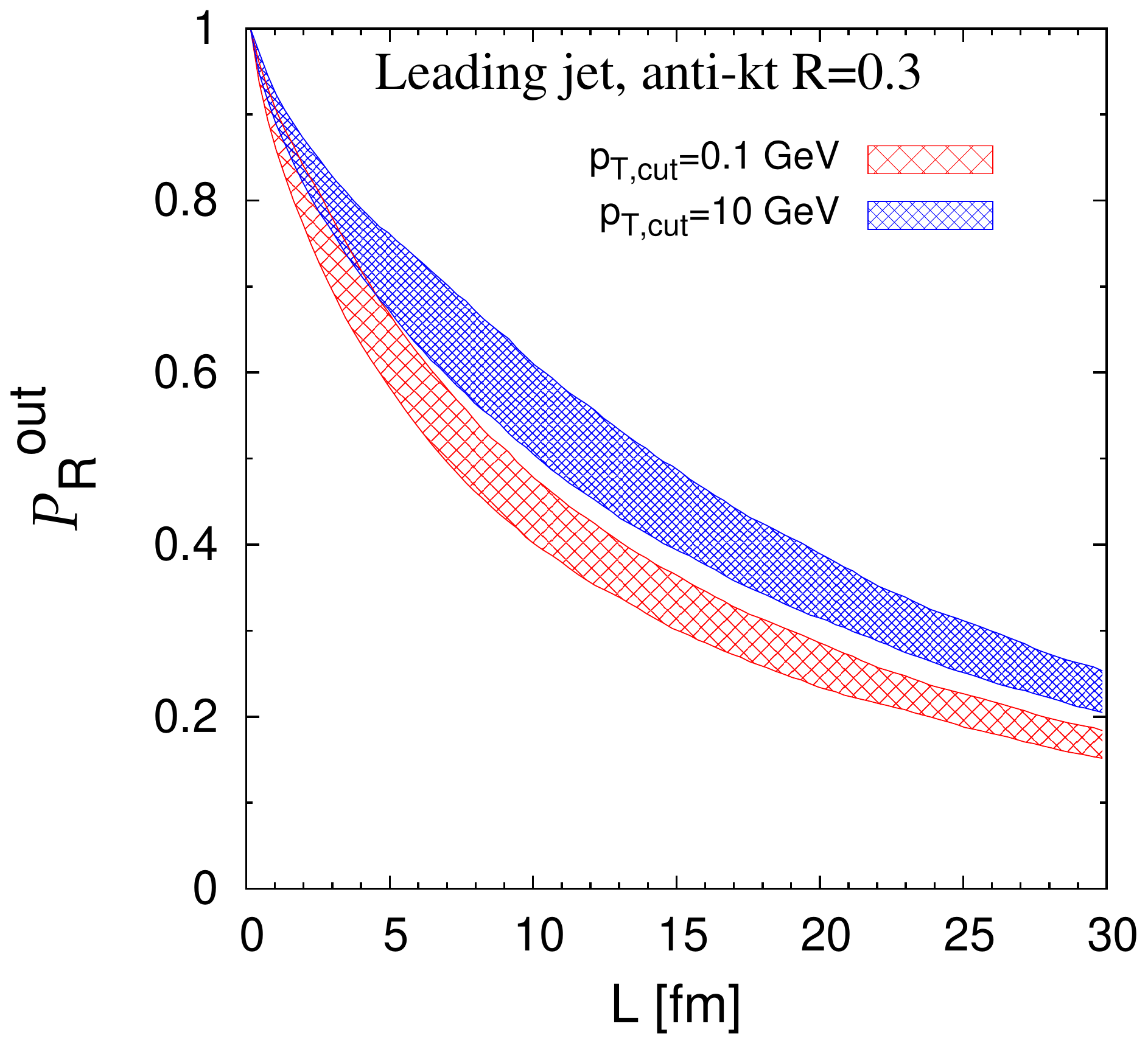}
\includegraphics[angle=0,width=.49\linewidth]{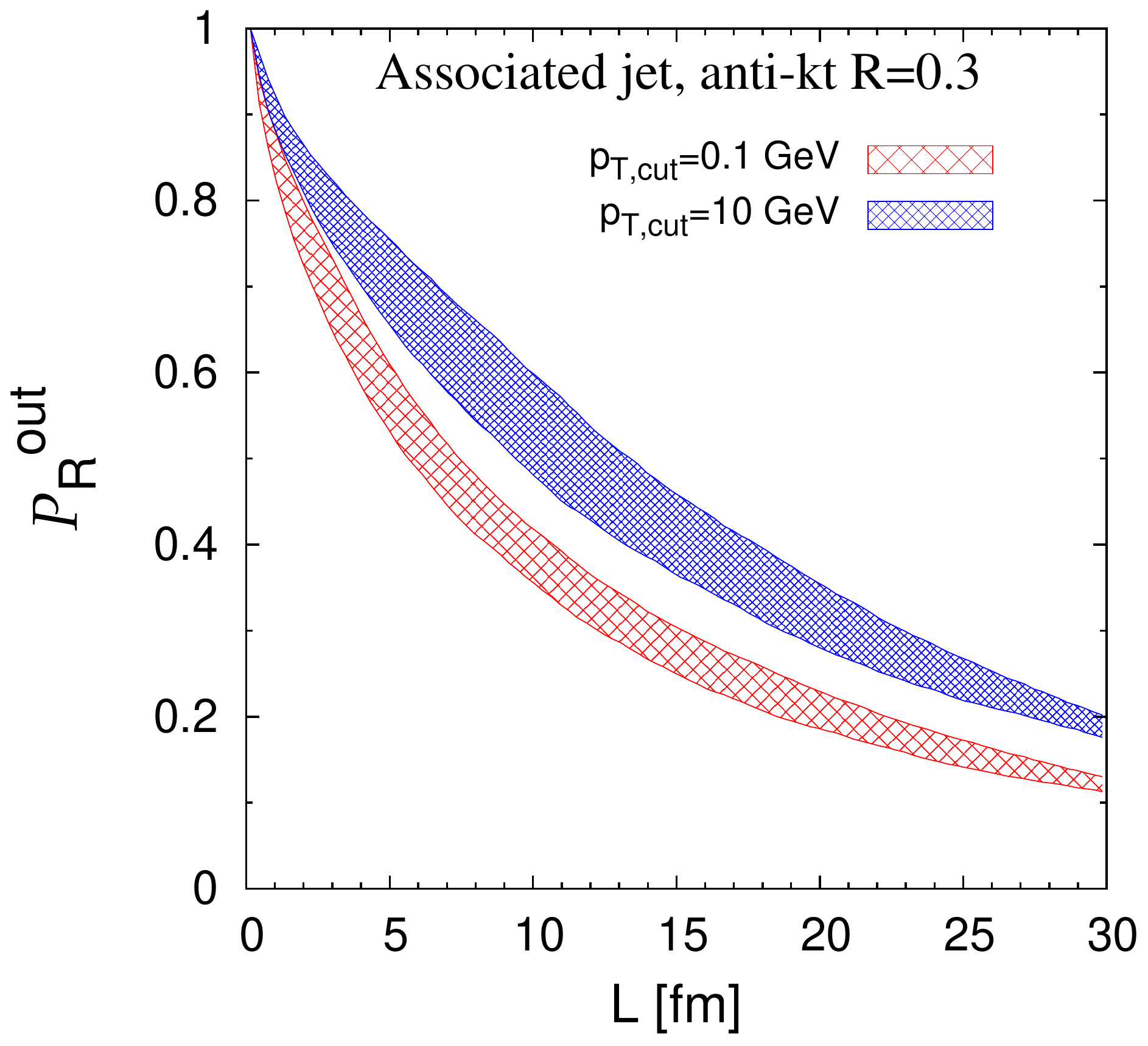}
\caption{\label{pouts}
Probability that the splittings of the parent partons occur at a time larger than $L$ in the frame where the parton is transverse for leading (left) and associated (right) jets in di-jet events of $p_{T,leading}> 100$ GeV and  $p_{T,associated}> 40$ GeV. The bands correspond to the two different estimates of the splitting time described in the text. A transverse momentum $p_T$-cut
at reconstrunction level has been introduced. As the $p_T$-cut is increased, late fragmentation patterns are favored.}
\end{figure}

From this sample we compute the restricted probability $ \mathcal{P}^{out}_R(L)$ for both the leading and associated jet, see Fig. \ref{pouts}. In these plots the bands correspond to the 
two models for computing the emission time described above. A transverse momentum cut of the gluons used in the jet
reconstruction has been introduced: in both plots the upper band corresponds to a $p_T$-cut of $10$ GeV and the lower 
one to $0.1$ GeV. For either cut, both in the leading and in the associated jet we observe that the shower develops in a 
rather long time. In fact, 
the distribution possesses a long tail at very large times, which reflects the logarithmic divergence in $Q^2$ of the splitting
kernel.  Remarkably, the probability of splittings occurring at times as large as  $L=5$ fm, comparable to the radius of a large nucleus, is larger than 50\%.

\section{\label{geometry}In-medium length in heavy ion collisions}
The dense hadronic matter produced in a heavy ion collision can probe and modify the time structure of the evolution process described above. 
The typical size of the medium produced by these collisions is of the order of the nuclear radius and, as we have seen, comparable to the characteristic time for the development of the vacuum shower. 
For this reason, we would like to compare the vacuum fragmentation pattern with the time extent that jets travel through the medium. It is clear that 
not all jets in a nuclear collision traverse the same amount of medium, as they are not all produced at the same point within the colliding region, and path length fluctuations must be taken into account. 
 These originate from simple geometrical considerations which we describe below.

As it is well known, the emission points of hard jets in the transverse plane $(x_0,y_0)$ are distributed according to the number of collisions per unit area, $T_{AA}(x_0,y_0,b)$
\be
T_{AA} (x_0,y_0,b)=T_A(x-b/2,y)T_A(x+b/2,y)
\ee
where the nuclear density profile $T_A(x,y)=\int dz \rho(x^2+y^2+z^2)$ is computed from the nuclear density $\rho(r)$, given by the standard Woods-Saxon potential \footnote{Since we will use this distribution only as an estimate, we will not take into account fluctuations in the 
distribution of nucleons in this computation.}. 
From these emission points, jets can travel at any direction in the transverse plane given by the (randomly selected) unit vector 
$\hat n=\left(\cos(\phi),\sin(\phi)\right)$. 

As the jet travels trough the medium, it traverses regions with varying density since there are  different number of participating nucleons at each point of the transverse plane. In addition, the medium formed in the collision is not static, but suffers a very strong longitudinal expansion which dilutes the system at later times. As customary, we will assume that the medium is boost invariant, so that all density variations occur in the transverse
plane. This assumption allows us to focus on the in-medium transverse path length, L, since we can always boost the system to a frame where the propagating parton is transverse to the beam. Note that this is also the frame where the distribution in
Fig. \ref{tau_f} is computed. To simplify the discussion, we will always refer to transverse lengths simply as lengths.

 Since medium effects are stronger when matter is denser, we introduce a ``density weighted" path length of the jet in the system given by
 \cite{Dainese:2004te,Eskola:2004cr}
\be
\label{ldef}
L=2 \frac{\int^\infty_0 d\tau \tau \rho(\x+\hat n \tau, \tau)}{\int^\infty_0 d\tau  \rho(\x+\hat n \tau, \tau)}
\, . 
\ee
where $\rho(\x,\tau)$ is the density in the (transverse) position $\x$ and time $\tau$.  Since we expect the bulk of the matter to scale with the number of participating nucleons in the collision, we will assume, as customary,  that the relevant density scale is proportional to the wounded nucleon profile $\rho_{WN}(\x)$, which we compute assuming $\sigma_{NN}=42$ mb. The explicit time dependence arises from the dilution of the system due to collective expansion. For simplicity, we will only take into account the longitudinal expansion in a Bjorken-like fashion, assuming a initialization time, $\tau_0=0.5$ fm.  We model the density, $\rho$, by 
\be
\label{Ldw}
\rho({\bf x},\tau)\propto \rho_{WN} ({\bf x}) \frac{\tau_0}{\tau+\tau_0} \,.
\ee  

\begin{figure}
\centering
\includegraphics[angle=0,width=.49\linewidth]{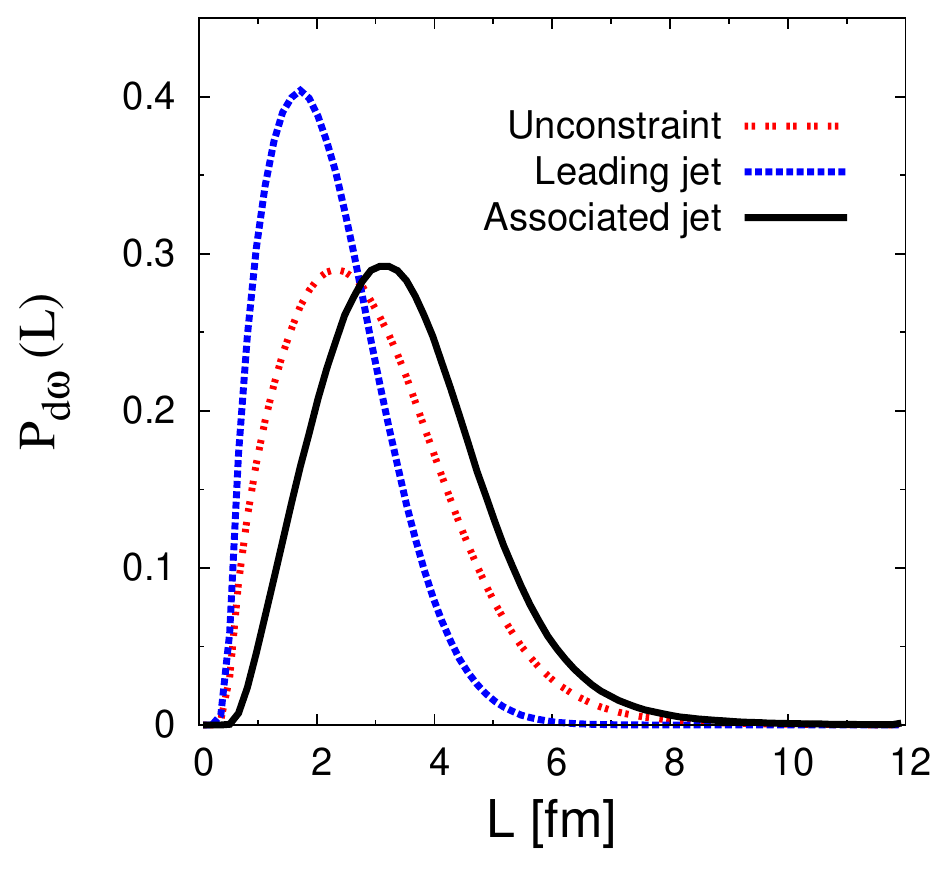}
\caption{\label{PdwL}
Path length distribution of high energy particles in a heavy ion collisions for a centrality class of 
$0-30$\%. The dashed line corresponds to neglecting energy loss effects. The dotted and solid lines
correspond to leading and associated jets, respectively, in di-jet events with energy loss effects described by the absorption model eq. \ref{absorpm}.
}
\end{figure}

 Averaging over
the transverse plane, over all possible directions of emission and over the impact parameter of the collision, we obtained the distribution 
of (density weighted) in-medium path lengths, $P_{dw}(L)$, which is  shown by the dashed line in Fig.~\ref{PdwL} for a centrality class 
of $0-30$\% (which coincides with the one used in the jet analysis \cite{CMS-analysis-summary} ).  Note that by taking into account the longitudinal expansion, as in eq. (\ref{ldef}) the distribution of effective path lengths is concentrated at smaller values than if the medium would have been taken as static \cite{Wicks:2005gt}, since at any point in the transverse plane the density drops with time.

 If medium effects were absent, or if it were possible to reconstruct all the initial energy of the jet (prior to medium interaction), the  path length distribution described above would coincide with the distribution probed by the finally observed jets. However, if those effects are not small, as inferred from the ATLAS \cite{Aad:2010bu} and CMS \cite{Chatrchyan:2011sx} measurements, a {\it surface bias} occurs, since the combination of energy loss and a steeply falling spectrum favors small in medium path lengths. The exact description of this effect demands a good microscopic description of the process of jet energy loss and its subsequent evolution, which is beyond the scope of this note, but a  simple estimate of this effect can be achieved by assuming an absorption model for the in medium jets \footnote{
 By absorption we do not mean that the jet disappears but that the energy shift is sufficiently large to make its contribution to the spectrum negligible.}. Motivated by the fact that in-medium radiative energy loss grows like $L^2$, we assume a survival probability for the jet given by
 \be
 \label{absorpm}
 A(L)=e^{-L^2/L^2_c}\, ,
 \ee
with $L_c$, the critical length, a model parameter.  The model is fixed by demanding that after convoluting with the initial geometry, the overall survival probability of an inclusive jet coincides with the jet $R_{AA}$. Motivated by preliminary $R_{cp}$ measurements for jets \cite{Angerami:2011is}, we assume $R_{AA}=0.5$, which yields $L_c=3.3$ fm. 

The trigger bias we have just described  leads to a different path length distribution probed by inclusive jets and for di-jets, since the requirement of a leading jet fixes the production point and forces the associated jet to travel  a longer distance; neglecting acoplanarity effects between the two jets, this can be obtained via eq.~(\ref{Ldw}) after substituting $\phi\rightarrow \phi+\pi$ in the definition of $\hat n$ and demanding that the leading jet is not absorbed. This distribution is shown by the solid line in Fig. \ref{PdwL} which, as expected, is shifted towards larger values.  However, both because of the large centrality bin that we have considered and the strong longitudinal expansion, most of the jets traverse an in-medium path length of less than $L=5$ fm.

\section{\label{foutside}Fragmentation outside of the medium}
The combination of the probability distribution in Fig. \ref{pouts} and the in-medium path length distribution Fig. \ref{PdwL} shows
that in the absence of medium-induced modifications of the branching process, and thus with jet evolution in matter proceeding as in the vacuum, a large fraction of the branchings would occur outside of the medium. 
The interactions between high energy partons and the medium affect the properties of  final jet observables; in particular,  additional medium-induced gluon radiation is expected to occur thus modifying the emission process. Clearly, the dynamics of those splittings that occur outside the 
medium remain unchanged, up to an overall reduction of the parent parton energy when it leaves the 
medium due to energy loss  \footnote{Note that color flow changes could alter the fragmentation pattern even for splittings outside of the medium \cite{Beraudo:2011bh}.}. Both this phase space reduction and the additional probability of splitting tend to accelerate the evolution process as compared to the vacuum. 
As a consequence, the in-medium formation time is modified, as it has been explicitly shown in the  in the context of radiative energy loss in \cite{Zapp:2008af,Zapp:2011ya}.
Thus, by convoluting the vacuum distribution
fig. \ref{pouts} and fig. \ref{PdwL}, we can establish an upper bound on the fraction of splittings $f_{out}$
that occur outside the medium
\be
f_{out}=\int dx P_{dw} \left(x\right) \mathcal{P}_R^{out}(x) \, .
\ee

\begin{figure}
\centering
\includegraphics[angle=0,width=.49\linewidth]{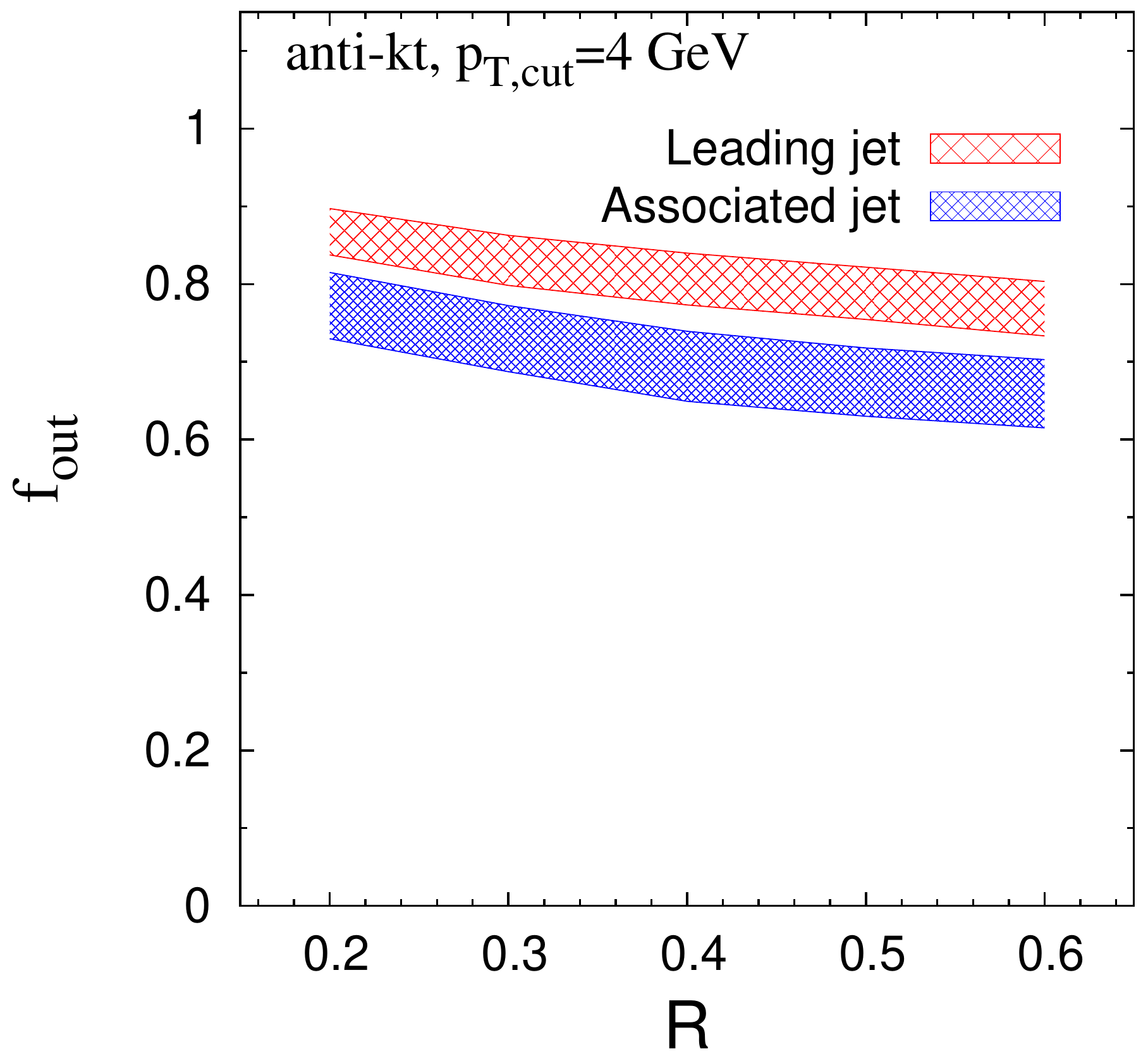}
\includegraphics[angle=0,width=.49\linewidth]{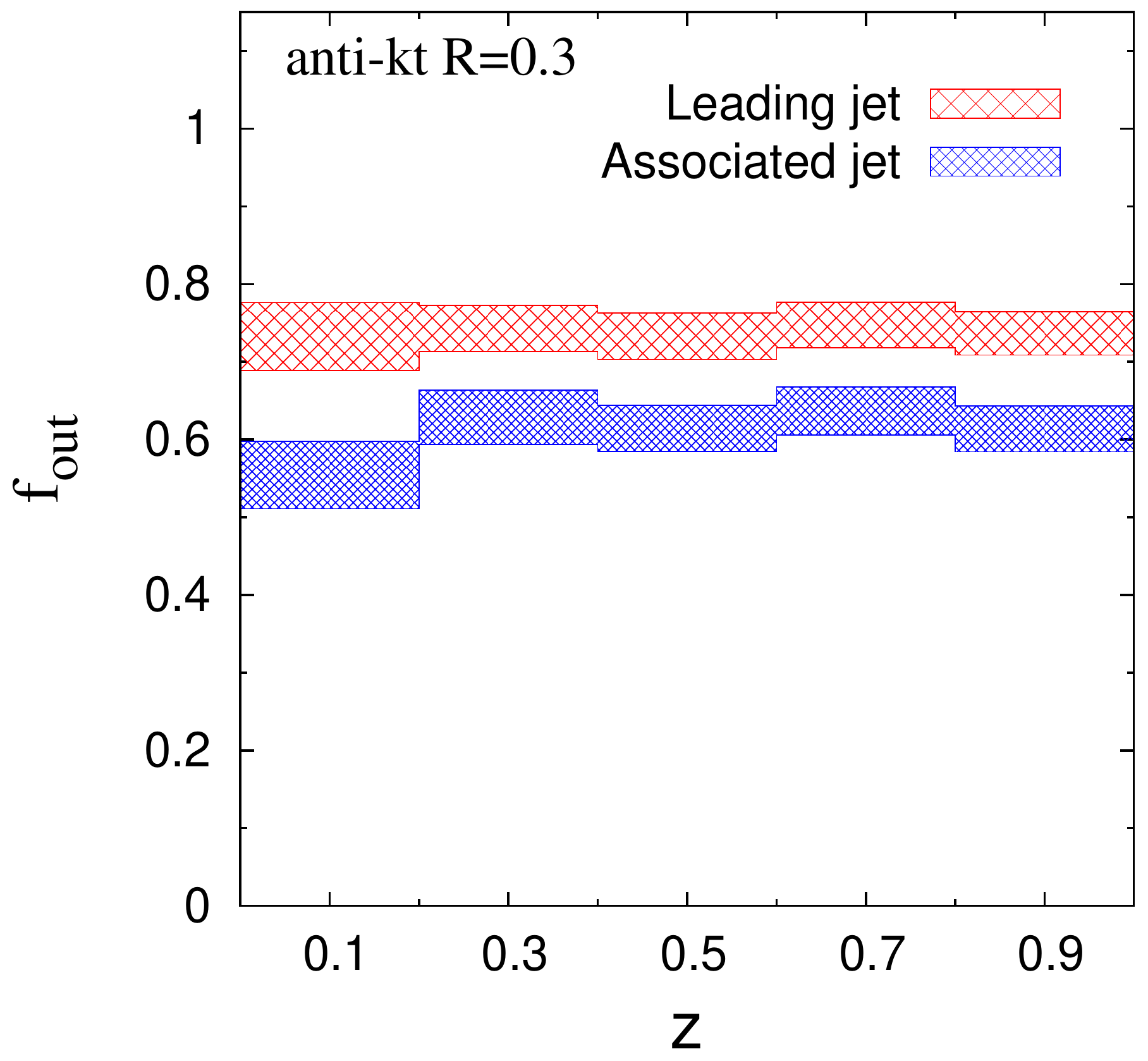}
\caption{\label{Convolution}
Fraction of splittings of the parent parton that occur outside of the medium, assuming that in heavy ion collisions jet evolution proceeds as in vacuum. The analysis is performed for di-jet events of leading jet transverse momentum $p_{T,leading}>100$ GeV and associated jet momentum $p_{T,associated}> 40$ GeV and for a lead-lead collision with centrality class $0-30$ \%. A very mild dependence of this fraction 
is observed both in the reconstruction radius $R$ (left) and in the $z-$fraction of transverse momentum (right). Note that in the two plots sample differently the distribution of splittings (see text for discussion).
}
\end{figure}

In Fig. \ref{Convolution} (a) we show $f_{out}$ as a function of the anti-kt radius R for both leading (upper band) and associated jets (lower band) with the energy and centrality cuts  as specified in previous sections. 
As for the computation of $\mathcal{P}^{out}_R$, the shown bands reflect the spread in the result due to the two different ways of estimating the time evolution described in section \ref{vactime}. In both cases, motivated by the the experimental analysis of jet distributions \cite{Collaboration:2011nsb},  
we have added a lower
$p_T=4$ GeV cut in the transverse momentum of the partons at reconstruction level. 
 From this plot we conclude that for 
leading jets most of the branching process occurs outside of the medium, with $f_{out}>80$\%. For the associated jet, where the in-medium path lengths are larger, there is a reduction in the fraction of outside splittings; however, even 
for those $f_{out}>70$\%. Additionally, our study  shows a very small dependence of $f_{out}$ on the reconstruction radius; small radii, 
which favor collinear fragmentation patterns, have a larger contribution of splittings outside of the medium in both 
cases, but the dependence is mild and even for radius as large as  $R=0.6$ the fragmentation is dominated by late splittings.

In Fig. \ref{Convolution} (b) we show a fragmentation function-like distribution in which the fraction of events
is computed in bins of the fractional transverse momentum of the parton in the jet 
\be
\label{zdef}
z=\frac{p_{T,parton}}{p_{T,jet}}\, ,
\ee
with $p_{T, jet}$ the modulus of the vectorial sum of transverse momenta of all the partons within the jet for a fixed
jet radius $R=0.3$. Once again, we observe, at best,  a very mild dependence on the energy of the fragment 
in both leading and associated jets; in both cases,  the majority of fragments occur outside
of the medium for all energies. Note that in this distribution $f_{out}$ is smaller, for all $z$-bins, than the corresponding value in 
Fig.  \ref{Convolution} (a). This is a consequence of our procedure to avoid double counting of splittings, since in the more 
inclusive distribution, Fig.  \ref{Convolution} (a), more partons share common splittings than in the more differential one  
Fig.  \ref{Convolution} (b). Since common splittings occur earlier in the evolution and they are only counted once, the 
sample used in Fig.  \ref{Convolution} (b) gives a larger weight to late splittings. The difference between the two procedures is 
not large and does not affect significantly the conclusions of our study. It should be kept in mind that our observations
are not based on a well grounded theoretical model and should not be taken for more than indication of the underlaying dynamics.

In addition to  this uncertainty, we would like to stress once more that our analysis has been performed at parton level 
and hadronization effects have not been taken into account. While the time scale estimation of the gluon shower is not affected by such effects, hadronization can alter the distribution of particles within the reconstruction cone, specially at small radius \cite{Soyez:2010rg}.  Furthermore, the decay of the final partonic system into hadrons will lead to a softer distribution of fragments at hadronic level than at partonic level. In this sense, the $p_T$-cut we have imposed in the distribution in Fig. \ref{Convolution} (a) is too low for a
comparison with experimental data; generically, hadrons with $p_T> 4$  GeV will be the result of the hadronization of more energetic partons. However, as we have shown in Fig. \ref{pouts}, imposing a higher $p_T$ cut pushes the distribution of splittings to later times, increasing the fraction $f_{out}$.

Finally, the particular value of $f_{out}$ we have obtained depends on the absorption model for jets that we have described in section \ref{geometry}. As stressed there, while we view this model as theoretically motivated, the exact length dependence of the jet energy loss is not known \cite{Liao:2008dk,Marquet:2009eq,Jia:2011pi}. A weaker length dependence than the one we have used would reduce the difference of path lengths between leading and associated jets and the value of $f_{out}$ for each of them would be between the two bands in Fig. \ref{Convolution}. A stronger length dependence would increase that separation, increasing $f_{out}$ for the leading jets and reducing it for the associated ones. However, 
unless a very strong dependence on $L$ is advocated, which is neither theoretical nor phenomenologically justified, we do not expect a dramatic change in the $f_{out}$ value.  In particular, for leading jets and for any length dependence, the fraction of relevant splittings outside of the medium will be always larger than $70$\% (in the kinematic range of interest and for $R=0.3$).

\section{\label{discussion}Summary and Discussion}
In this note we have studied the time structure of jet evolution in the vacuum  by following the 
DGLAP chain and assigning to each splitting a decay time given by
  eq. (\ref{tau_f}).
  For the kinematic range currently explored in jet measurements at the LHC, several splittings occur
as the partons produced in an elementary collision shower towards final hadrons. We have studied
the time distribution of those splittings from the parent parton, fig. \ref{pouts},
defined as the latest step in the decay chain common to all the fragments within the reconstructed cone, 
and concluded that for those jets
the fragmentation process in vacuum takes a long time, comparable to the nuclear radius. 
We have also observed that this time is larger than the typical path lengths 
travelled by jets in heavy ion collisions, fig. \ref{PdwL}. By convoluting these two distributions
we found that more than 80\% (70 \%) of the splittings in leading (associated) jets occur at a 
time greater than the path length of an equivalent parton propagating in the medium created in a heavy ion collision.

The fact that the branching process takes such a long time seems to be in agreement with 
preliminary observation by both ATLAS \cite{Angerami:2011is} and CMS \cite{Collaboration:2011nsb}, which have shown that 
for di-jet samples with leading jet
of $E_T$ greater than $100$ GeV  (ATLAS and CMS)  and a associated jet of $E_T$ greater than 40 GeV (CMS),
and with relatively small anti-$k_T$ reconstruction radius, there is no observed modification of the 
fragmentation pattern. Surprisingly, this is so even for associated jets in di-jet events
with a large transverse energy imbalance,
for which a strong distortion of the asymmetry distribution has been observed.
While the average energy of  associated jets is reduced, their fragmentation function is indistinguishable
from that of associated jets in p-p collisions at the same (associated) jet energy.  

A natural
explanation for this agreement is that most of the relevant part of the jet evolution happens at late times, 
outside of the medium, where the branching process cannot be altered by its interaction with the medium.
This is precisely what we have observed for vacuum jets, since most of the splittings of the 
parent partons occur late. 
Our study supports this explanation: if  vacuum branching occurs at very early times, it must be 
affected by the medium and, then, it is hard to understand how such modification could leave the 
fragmentation functions unchanged, unless the medium effects were small.  
However, the strong distortion of the di-jet asymmetry distribution in a heavy-ion environment reported in 
\cite{Aad:2010bu,Chatrchyan:2011sx} indicates that this is not the case.
Thus, while we cannot rule out that modified branching patterns may lead to the same fragmentation distribution, we will adopt
the more natural assumption that current fragmentation function measurements are mostly
sensitive to the dynamics outside of the medium.

Since medium effects are strong, the in-medium distribution of splittings may be altered and conclusions 
based on a vacuum physics analysis may not be valid. A fair comparison to data demands a
dynamical understanding of the jet-medium interactions. However,  the main mechanism responsible 
for the observed energy loss of high energy partons remains unclear and several models are 
available. Enlightened by  our study,
we discuss the effect of three different sets of those models for the splitting distribution:
\begin{enumerate}
\item The interaction with matter may modify significantly the radiation 
of (almost collinear) partons, increasing the rate of radiation and leading to a large energy loss \cite{Borghini:2005em,Armesto:2009zc,Armesto:2009ab,Renk:2010zx}. In this case, 
the evolution process becomes faster than in the vacuum, since increasing the radiation rate would lead 
 to a faster evolution, as compared to the vacuum, and most of the virtuality would be relaxed in the medium. In this case
 the fragmentation functions should be sensitive to this modified pattern and show a modification with respect to the 
 vacuum. 
 \item The medium could add extra relatively hard induced radiation via scattering with the matter constituents without significantly 
 modifying the relaxation of virtuality. This corresponds to the usual mechanism of radiative energy loss \cite{Wiedemann:2009sh,d'Enterria:2009am,CasalderreySolana:2007zz,Majumder:2010qh,Jacobs:2004qv,Gyulassy:2003mc}. Since the 
 rate of evolution is not modified, a significant part of the branchings of the parent parton would take place still outside of the medium.
 Energy loss is due to additional early splittings of the parent parton, which only affect the late splittings by reducing the available 
 energy.  The induced splittings add extra fragments into the final hadron distribution but with an angular pattern which is different from
 vacuum emissions.
  \item Energy degradation could be due to many small losses transported to large angles. This is the case for collisional energy loss \cite{Wicks:2005gt}, 
 coupling to collective modes \cite{CasalderreySolana:2004qm},  
  transport of soft quanta at large angles \cite{CasalderreySolana:2010eh}
 or soft large angle radiation via interference effects \cite{MehtarTani:2010ma,MehtarTani:2011tz,Armesto:2011ir}.  As in the previous case, 
  energy loss would occur at early times, reducing the energy of the parent parton, but leaving the late branchings, which influence the most the final
 distribution of fragments, unmodified. For all these models, the 
 contribution of the lost energy to the fragmentation function would have a weak dependence on the reconstruction radius
 $R$. Additionally,  in all these scenarios the energy is degradated into modes with momenta
  of the same order as the underlying event, which would hide their contribution to the fragmentation function. 
 \end{enumerate}
 
 Preliminary data from both CMS \cite{Collaboration:2011nsb} and ATLAS \cite{Angerami:2011is} clearly disfavor the first scenario, since no modification of the fragmentation 
 function has been observed. However, from the available data it is not possible to favor any of the two last 
 scenarios: if the additional radiation by the second mechanism occurs at sufficiently large angles, its effect in 
 small radius cones is negligible.  Clearly, extending these measurements, 
 specially for the associated jet, to larger $R$ values could allow us to distinguish among those two possibilities; however, the large
 fluctuating backgrounds complicate 
 these measurements. Fortunately, there are other qualitative features which could 
 help disentangle the correct mechanism. 
 Since, as we have argued, the extra relatively hard radiation of the second scenario must lye
 outside of the jet cone,  it leads to an increase 
 in the number of (semi-hard) jets associated to a parton that traverses the medium. Assuming a sufficiently strong surface bias,
 trigger jet events in heavy ion collisions would show a larger number of softer jets in the away side, specially for those 
 events with large di-jet asymmetries. On the contrary, such an effect would be much milder for the third scenario 
 for which the jets would be accompanied by an increase of the underlying event activity as compared to the heavy ion background.

To conclude, we would like to remark that while the typical scale of the vacuum distribution in fig. \ref{pouts}
is larger than those probed by the medium, it is not asymptotically large. Thus, in a heavy ion collision, 
a non-negligible part of the evolution takes place within the medium. However, to the best of our knowledge up to date there is not a theoretically satisfactory description of the space-time structure of the vacuum fragmentation process, which is a fundamental building block for the study of medium modification of jet physics. With the estimates presented in this note we have explicitly showed the importance of the understanding of this structure for the description of jet data in heavy ion collisions.

\section{Acknowledgments.}
We thank M. Cacciari for his valuable comments on the manuscript. We also thank N. Armesto, C. Marquet,  S. Sapeta, G. Salam and U. Wiedemann for useful discussions.  
The work of JCS is supported by a Ram\'on y Cajal fellowship. The work of PQA is supported by the French ANR under contract ANR-09-BLAN-0060.

\end{document}